\documentclass[preprint,aps,nofootinbib]{revtex4}

\usepackage{graphicx}

\usepackage{amsmath}
\usepackage{amsfonts}
\usepackage{amssymb}

\newcommand{\be}{\begin{equation}}
\newcommand{\ee}{\end{equation}}
\newcommand{\bear}{\begin{eqnarray}}
\newcommand{\eear}{\end{eqnarray}}
\newcommand{\beqstar}{\begin{eqnarray*}}
\newcommand{\eeqstar}{\end{eqnarray*}}

\begin{document}


\vspace*{1.5cm}
\title{Top Partners in Little Higgs Theories with $T$-parity}

\author{Hsin-Chia Cheng$^{a,b}$,  Ian Low$^c$ and Lian-Tao Wang$^a$}
\vspace*{.4cm}
\affiliation{
\vspace*{0.4cm}
$^a$Jefferson Physical Laboratory,
Harvard University, Cambridge, MA 02138 \\
$^b$Department of Physics, University of California, Davis, CA 95616 \\
$^c$School of Natural Sciences, Institute for Advanced Study, Princeton, NJ 08540
\vspace*{0.5cm}}

\begin{abstract}
\vspace*{0.5cm} We consider a class of little Higgs theories with
$T$-parity where {\it all\/} new particles responsible for
canceling the standard model contributions to the one-loop
quadratic divergences in the Higgs potential are odd under
$T$-parity, including the heavy top partner which was previously
taken to be $T$-even. The new construction significantly
simplifies the spectrum in the top sector and completely changes
the phenomenology of the top partner. At hadron colliders the
signals of this class of $T$-invariant models appear to be even more
similar to supersymmetry. We initiate a study on the collider
phenomenology and discuss the challenge of distinguishing this
class of models from supersymmetry at the Large Hadron Collider.

\end{abstract}


\maketitle

\section{Introduction}
\label{sec:introduction}

In the past few years a new approach to the electroweak symmetry
breaking which is also weakly-coupled at the TeV scale, dubbed
little Higgs theories, was proposed in
Ref.~\cite{Arkani-Hamed:2001nc}. In these theories the Higgs field
arises as a pseudo-Nambu-Goldstone boson (PNGB). The low energy
effective theory is described by a non-linear sigma model with the
peculiar property that the quadratically divergent contributions
to the Higgs mass parameter are absent at one loop. This allows
the cutoff of the effective theory to be raised to $\sim10$ TeV
without introducing fine-tunings in the Higgs potential. A cutoff
at 10 TeV, on the other hand, is just high enough to suppress the
non-calculable higher dimensional operators at the cutoff scale
\cite{Barbieri:1999tm, Han:2004az} in order to avoid the ``little
hierarchy problem.''

There are several different types of little Higgs theories
\cite{Arkani-Hamed:2002qx,
Arkani-Hamed:2002qy,Low:2002ws,Kaplan:2003uc,Chang:2003un,
Chang:2003zn,Skiba:2003yf,Schmaltz:2004de,Cheng:2003ju,Cheng:2004yc,
Low:2004xc}, as well as some examples of UV extensions above 10
TeV \cite{Katz:2003sn,Kaplan:2004cr,Thaler:2005en,Thaler:2005kr},
and extensive phenomenological studies have been performed
\cite{Arkani-Hamed:2002pa,Burdman:2002ns,Han:2003wu,Hewett:2002px,
Csaki:2002qg,Csaki:2003si,
Gregoire:2003kr,Kilic:2003mq,Chen:2003fm,Kilian:2003xt,Perelstein:2003wd,
Azuelos:2004dm,Hubisz:2004ft,Marandella:2005wd, Hubisz:2005tx,
Han:2005dz, Han:2005ru}. Despite many variations, the generic
spectrum
of little Higgs theories has new scalars, new fermions, and new 
gauge bosons at around 1 TeV to cancel the quadratically divergent
contributions to the Higgs mass-squared which come from the
standard model particles. The contributions of these new particles
to the electroweak observables, which are calculable within the
low-energy effective theory, turn out to be quite model-dependent.
Generically tree-level corrections to the the electroweak
observables would require raising the mass of the new particles to
be much higher than 1 TeV, thus re-introducing fine-tunings in the
Higgs potential~\cite{Marandella:2005wd}. However, it was realized
that these tree-level contributions can be eliminated in a
systematic way in many little Higgs models by introducing a
$T$-parity, under which standard model particles are even and the
new heavy scalars and gauge bosons are
odd~\cite{Cheng:2003ju,Cheng:2004yc, Low:2004xc}. The leading
corrections to electroweak observables are then loop suppressed so
that models with $T$-parity can be made consistent with the
electroweak constraints without raising the mass of the new
particles.

However, in the previous constructions of $T$-parity, the top
sector is quite complicated. There are more than one new colored
fermions at or below the TeV scale with both even and odd
$T$-parities. In particular the heavy top partner which cancels
the top-quark quadratic divergence is $T$-even. If this assignment
is inevitable, then the new top partner can be a smoking gun in
distinguishing little Higgs theories from other scenarios such as
$R$-parity conserving supersymmetry (SUSY), and previous studies
on the phenomenology in the top sector of little Higgs theories
without $T$-parity, which rely on the single production of the
heavy top partner \cite{Perelstein:2003wd,Azuelos:2004dm,
Han:2005ru}, can still be applied to models with $T$-parity. In
this paper we show that in fact there is a way to implement the
$T$-parity in the top sector such that only one top partner is
required to be $\lesssim$ TeV and it is odd under $T$-parity; the
spectrum in the top sector is simplified. As a result, every
standard model contribution to the one-loop quadratic divergence
of the Higgs mass is canceled by that of a new particle with the
same spin but opposite $T$-parity.

The introduction of $T$-parity has drastic effects on the
phenomenology of little Higgs theories
\cite{Cheng:2004yc,Hubisz:2004ft, Hubisz:2005tx,Dib:2005re}. The $T$-odd
particles have to be pair-produced in collider experiments and the
lightest $T$-odd particle (LTP) is stable. If the LTP is neutral,
it can be a good dark matter candidate and also gives rise to
missing energy signals in colliders. Now with the $T$-odd top
partner, the searches for the top partner based on single
productions as well as decays into $Zt$, $Wb$ and
$h\,t$~\cite{Perelstein:2003wd,Azuelos:2004dm} are no longer
valid. The $T$-odd top partners now also need to be pair-produced
and their decays to LTP's give rise to missing energies plus jets,
similar to the decays of top squarks in SUSY. One can see that
$T$-parity in little Higgs theories plays a similar role as the
$R$-parity in supersymmetric theories and the $KK$-parity in
Universal Extra Dimensions
(UEDs)~\cite{Appelquist:2000nn,Cheng:2002iz,Cheng:2002ab}. This
similarity makes it a serious challenge to distinguish these
scenarios at a hadron collider if such signals are discovered, as
the spin information of the new particles is difficult to obtain.

In the following we concentrate on the top sector of the little
Higgs theories. In the next section we discuss the implementation
of $T$-parity with a $T$-odd heavy top partner for generic little
Higgs theories. In section III we discuss the impact of a $T$-odd
top partner on the electroweak corrections. In section IV we study
the collider phenomenology of the $T$-odd top partner at the Large
Hadron Collider (LHC). Then we conclude, which is followed by an
appendix exhibiting how to implement our construction in two
popular examples of little Higgs theories.

\section{$T$-odd heavy top partners}
\label{toppartner}

The central idea of the little Higgs mechanism is the ``collective
symmetry breaking:'' no single interaction in the lagrangian
breaks enough global symmetries under which the Higgs particle is
an exact Nambu-Goldstone boson; only in the presence of at least
two interactions does the Higgs become a PNGB and acquire a mass
term suppressed by the product of the two coupling constants in
front of the two interactions involved. If one can make sure that,
under naive dimensional analysis \cite{Manohar:1983md}, each
coupling comes with a loop factor, then the leading quadratic
divergence in the Higgs mass can only arise at the two-loop order.
In actual model-building the collective breaking is implemented in
the scalar, the gauge, and the top sector separately, since each
particle in the same representation of a global symmetry must have
the same spin. This is why the standard model contribution in each
sector is canceled by that of a new particle with the same spin.
There are already many different types of little Higgs models in
the literature and the way to implement $T$-parity in models based
on symmetric spaces has been discussed in
Refs.~\cite{Cheng:2003ju,Cheng:2004yc, Low:2004xc}. In this paper
we focus on the top sector and discuss a new and simpler way to
implement the $T$-parity and its phenomenological consequences.

First let us review how the collective breaking in the top sector is achieved.
To be specific, we illustrate it with $SU(3)$ global symmetries.
A vector-like pair
of colored Weyl fermions, $U$ and $U^c$, that are singlets under $SU(2)_W$ are
introduced and couple to the Higgs scalar and the third generation quarks
$q_3=(d_3,\,u_3),\, u_3^{c\prime}$ through interactions
\begin{equation}
\label{top1}
{\cal L}_t =  \lambda_1 f\, (d_3,\, u_3,\, U)\, V
\left( \begin{array}{c} 0 \\ 0 \\ u_3^{c \prime} \end{array}\right) +
\lambda_2 f\, U^c U
  + {\rm h.c.} \quad ,
\end{equation}
where $V=V(\Sigma)$ is a function of the Goldstone
bosons $\Sigma=\exp(2i \pi/f)$. Under $SU(2)_W$ the Goldstone bosons are
decomposed into a Higgs doublet $H$, a triplet $\Phi$ and a singlet $S$,
\begin{equation}
\pi \sim \frac{1}{\sqrt{2}}
\left( \begin{array}{cc} \Phi+S/\sqrt{6} & H \\ H^\dagger & -2S/\sqrt{6} \end{array}\right).
\end{equation}
Each $\lambda_i$ interaction individually preserves enough global
symmetry to keep the Higgs massless; only when both $\lambda_1$
and $\lambda_2$ are present  does the Higgs acquire a mass. Here
we follow the convention in Ref.~\cite{Perelstein:2003wd} to use
letters $u$ or $U$ for weak eigenstates and $t$ or $T$ for mass
eigenstates. When expanding ${\cal L}_t$ to the first order in the
Higgs doublet $H$, we have
\begin{equation}
{\cal L}_t  = \lambda_1 u_3^{c \prime} \left( f U+{\sqrt{2}} H q_3 \right)
+ \lambda_2 f U^c U + \cdots \quad .
\end{equation}
We see that a linear combination of $u_3^{c \prime}$
and $U^c$ marries $U$ to become massive at $\sim \lambda f$,
whereas the orthogonal combination remains massless and has a
Yukawa coupling to the Higgs doublet. To check the cancellation of
one-loop quadratic divergence explicitly, one needs to go to the
quadratic order in the Higgs doublet
\cite{Perelstein:2003wd,Han:2005ru}. Writing in terms of mass
eigenstates,
\begin{eqnarray}
\label{mass1}
t&=& u_3, \quad \quad t^c = \frac{\lambda_2 u_3^{c \prime} - \lambda_1 U^c}
                                                       {\sqrt{\lambda_1^2+\lambda_2^2} },  \\
\label{mass2}
T&=& U, \quad \quad T^c =   \frac{\lambda_1 u_3^{c \prime} + \lambda_2 U^c}
                                                       {\sqrt{\lambda_1^2+\lambda_2^2}  }   ,
\end{eqnarray}
the lagrangian now becomes
\begin{equation}
\label{lt}
{\cal L}_t = m_T T^c T+ {\lambda_t} h t^c t +
        {\lambda_T} h T^c t  -
          \frac{\lambda_T^\prime}{2 m_T} h^2 T^c T + \cdots \quad,
\end{equation}
where
\begin{figure}[t]
\includegraphics[scale=0.75,angle=0]{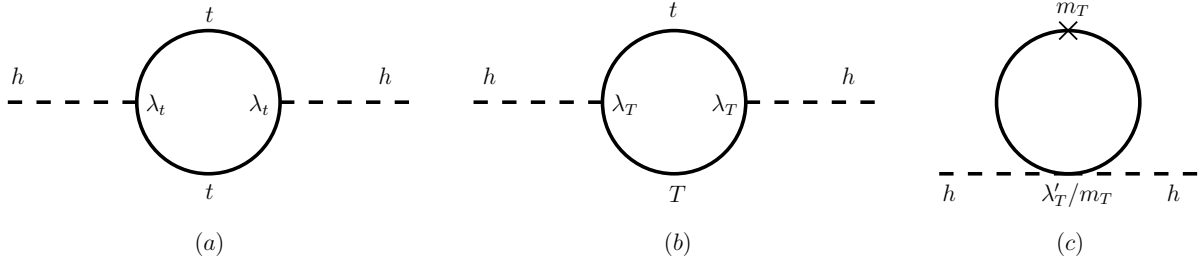}
\caption{\label{cancellation}{\it Cancellation of top quadratic divergences.}}
\end{figure}
\begin{equation}
\label{parameters}
m_T = \sqrt{\lambda_1^2+\lambda_2^2}\,  f, \quad \quad
\lambda_t = \frac{\sqrt{2}\lambda_1\lambda_2}{\sqrt{\lambda_1^2+\lambda_2^2} } , \quad \quad
\lambda_T= \frac{\sqrt{2}\lambda_1^2}{\sqrt{\lambda_1^2+\lambda_2^2} }, \quad \quad
\lambda_T^\prime = 2\lambda_1^2,
\end{equation}
and $h$ is the neutral component in the Higgs doublet $H=(h^+\
h)^T$. There are three diagrams as shown in Fig.~\ref{cancellation} contributing to the
quadratic divergence of the Higgs mass and they involve $\lambda_t$,
$\lambda_T$, and $\lambda_T^\prime$, respectively. The
cancellation among the three contributions is guaranteed by the
relation
\begin{equation}
\label{cancel}
\lambda_T^\prime = \lambda_t^2 + \lambda_T^2.
\end{equation}
The relation provides an important test of the little Higgs mechanism.

Previous implementations of $T$-parity in the top sector simply
add to the lagrangian the images of the original interactions
under $T$-parity to make it $T$-invariant. Because of the $h T^c
t$ interaction involved in the diagram $(b)$ of
Fig.~\ref{cancellation}, the heavy top partner necessarily has the
same even $T$-parity as the top quark. In these constructions,
there are additional $T$-odd colored fermions introduced to avoid
large Higgs mass generated at higher
loops~\cite{Cheng:2003ju,Cheng:2004yc,Low:2004xc}.

Here we show that a simple modification of Eq.~(\ref{top1}) could make the
implementation of $T$-parity easier, and the quadratically divergent contribution to the Higgs mass from the top quark loop can in fact be canceled
by a $T$-odd top partner. We take the lagrangian of the top sector to be
\begin{equation}
\label{topnew}
{\cal L}_t =  \lambda f\, (d_3,\, u_3,\, T)\; V
\left( \begin{array}{c} 0 \\ 0 \\ U_a^{c} \end{array}\right)  +
                      \lambda f\, (d_3,\, u_3,\, -T)\; T[V]
\left( \begin{array}{c} 0 \\ 0 \\ U_b^{c} \end{array}\right)
+ {\rm h.c.} \quad ,
\end{equation}
where $T[V]\sim \Omega V^\dagger \Omega$, $\Omega= {\rm diag}
(1,1,-1)$, is the $T$-image of $V$. The two terms are simply
$T$-images of each other with $U_a^c$ mapped to $U_b^c$ and vice
versa. Each term preserves enough global symmetry which protects
the Higgs mass from being generated and the collective breaking is
realized.

Expanding the lagrangian (\ref{topnew}) to quadratic order in the
Higgs field we have
\begin{equation}
\label{newexpand} {\cal L}_t = \sqrt{2}\lambda f \, T\, T^c +
2\lambda\, h t t^c  - \frac{\sqrt{2}\lambda}{f} h^2 T\, T^c +
\cdots ,
\end{equation}
where
\begin{equation}
T^c= \frac{1}{\sqrt{2}}(U_a^c -U_b^c),\quad
t^c=\frac{1}{\sqrt{2}}(U_a^c+U_b^c)
\end{equation}
are odd and even under $T$-parity respectively. In terms of the parameters
in Eq.(\ref{lt}),
\begin{equation}
m_T = \sqrt{2} \lambda f, \quad \lambda_t = {2} \lambda, \quad
\lambda_T = 0, \quad \lambda_T^\prime= 4 \lambda^2.
\end{equation}
One can see that the relation ensuring the cancellation of top
quadratic divergences, Eq.~(\ref{cancel}), is satisfied, and the
divergence is canceled by a vector-pair of $T$-odd fermions.
Diagram $(b)$ of Fig.~\ref{cancellation} is now absent and the
divergence is canceled between diagrams $(a)$ and $(c)$.
In fact, this is similar to and inspired by how the top quadratic divergence is canceled
in the ``simple little Higgs'' model \cite{Kaplan:2003uc}.

In the above example, the one-loop quadratic divergence is
canceled by introducing an $SU(2)_W$ singlet top partner to
complete the left-handed top-bottom doublet into a triplet. One
can easily see that it is also possible to cancel the quadratic
divergence by completing the right-handed top quark into a
triplet. In that case, the new $T$-odd partner will be an
$SU(2)_W$ doublet instead. We will only consider the singlet case
for simplicity.

We would like to emphasize that the structure of the couplings
discussed here is quite general and model-independent. In fact,
this construction can be applied to every little Higgs theory in
the literature so far. For further demonstrations, we give
explicit constructions of the new top sectors for the popular
moose model and littlest Higgs model in the Appendix.

\section{Electroweak corrections}

So far we have concentrated on the top sector of the model. It will be useful
to summarize the features and generic spectrum of little Higgs theories with $T$-parity
before we go on to discuss the interactions of $T$-odd top partners, which are crucial
to understand the electroweak corrections, as well as the collider phenomenology.

Generic little Higgs theories have at around 1 TeV new particles responsible for
canceling the standard model quadratic divergences. They are a new set of heavy
$SU(2)\times U(1)$ gauge bosons $(W_H^{\pm}, W_H^0, B_H)$, new heavy
scalars $\Phi$, which could be triplet or singlet under $SU(2)_W$,
and new fermions $(T^c,T)$ that are partners of the top quark.
It is shown, in Refs.~\cite{Cheng:2003ju,Cheng:2004yc,Low:2004xc} and the previous
section, that these new particles can all be charged under the $T$-parity.
In addition to the new particles involved in the cancellation of
quadratic divergences, there is also a vector-like $T$-odd doublet
partner for every standard model fermionic doublet. (We will call
them the ``mirror fermions'' though they are vector-like and
contain both chiralities.) These mirror fermions serve to cut off
the contributions to the standard model four-fermion interactions
from the Goldstone boson loop~\cite{Low:2004xc,Hubisz:2005tx}. The
four-lepton interactions are now strongly constrained by the LEP
II data and the mirror leptons are required to be $\lesssim$ 2
TeV. The constraints on four-quark operators are weaker so the
mirror quarks can be much heavier. Such heavy states may not be
directly observed at the LHC, but they may induce higher
dimensional operators of the light states, suppressed by their
masses. So we include the effects of these mirror fermions in the
discussions below.

As emphasized already, the motivation for $T$-parity is to
eliminate leading order corrections to precision electroweak
observables in little Higgs models. Without $T$-parity the scale
of symmetry breaking $f$ would generally be required to be higher
than $3-4$ TeV which re-introduces fine-tunings in the Higgs
potential. With $T$-parity the electroweak corrections from
$T$-odd particles are only loop-induced. In
Refs.~\cite{Cheng:2004yc,Hubisz:2005tx} it is found that, in some
region of the parameter space for the $SU(5)/SO(5)$ model,
 $f$ only needs to be larger than 500 GeV in order
to satisfy the precision
electroweak measurements. The strongest constraints there come from
corrections to the
$\rho$ parameter and $Z \to b\bar{b}$ vertex due to
heavy gauge bosons as well as heavy top partners, but not from the triplet
scalars since they are $T$-odd and not allowed to have a nonzero vacuum
expectation value (VEV).

In the old way of implementing collective breaking in the top
sector, the leading electroweak corrections from the heavy top
partners originate from the $\lambda_T$ term in the effective
lagrangian, Eq.~(\ref{lt}), which gives a small mixing between $t$
and $T$ after the electroweak symmetry breaking when the Higgs
gets a VEV: $\langle h \rangle = v/\sqrt{2}$. After diagonalizing
the mass matrix, there is the physical top quark $t_L$ and a heavy
top partner $T_H$ \cite{Han:2003wu}
\begin{eqnarray}
t_L&=& c_L t - s_L T,
\phantom{aaaaaaaaa}\quad T_H = s_L t + c_L T\ ; \\
s_L &=& \frac{\lambda_T}{\sqrt{2}} \frac{v}{m_T} + {\cal O}(v^3/m_T^3) , \quad
  c_L= 1 -  \frac12 \left(\frac{\lambda_T}{\sqrt{2}} \frac{v}{m_T}\right)^2.
\end{eqnarray}
The important observation
at this point is that, since $(T,T^c)$ is a Dirac-pair of $SU(2)$ singlets,
at leading order the heavy top partner $T_H$ enters into
precision electroweak observables only through the mixture with $(t,t^c)$.
Explicit calculations for the littlest Higgs theories with $T$-parity
\cite{Cheng:2004yc,Hubisz:2005tx} showed that the corrections
to the $\rho$ parameter and $Z\to b\bar{b}$ vertex from the heavy top partners are,
assuming $m_T \gg m_t$ ,
\begin{eqnarray}
\delta \rho &=& \frac{3}{32\pi^2} \frac{e^2}{s_w^2 c_w^2}\frac{m_t^2}{m_Z^2} s_L^2
       \left[ \log\frac{m_T^2}{m_t^2} -1 +\frac12\left(\frac{\lambda_1}{\lambda_2}\right)^2
         \right ], \\
\delta g_L^{b\bar{b}} &=& \frac1{8\pi} \frac{\alpha\, g}{c_w s_w^2} \frac{m_t^2}{m_W^2}
 s_L^2 \log\frac{m_T^2}{m_t^2}, \label{zbb}
\end{eqnarray}
where $s_w^2=1-c_w^2=\sin^2\theta_w$ and $\theta_w$ is the Weinberg angle. Indeed,
both equations above are proportional to the mixing parameter $s_L$.

In our new implementation of $T$-parity in the top sector, there
is no mixing between $t$ and $T^c$ even after electroweak symmetry
breaking since $\lambda_T=0$ in Eq.~(\ref{newexpand}), and hence
$s_L=0$. Moreover, this result holds to all orders in $f$ and $v$
as long as $T$-parity remains unbroken since $(T,T^c)$ is odd
whereas $(t,t^c)$ is even. However, the $T$-odd top partners do
contribute to $\delta \rho$ and $\delta g_L^{b\bar{b}}$ at higher
orders, which generally involve the following four type of
vertices, with the magnitude of their coefficients indicated up
front,
\begin{figure}[t]
\includegraphics[scale=0.85,angle=0]{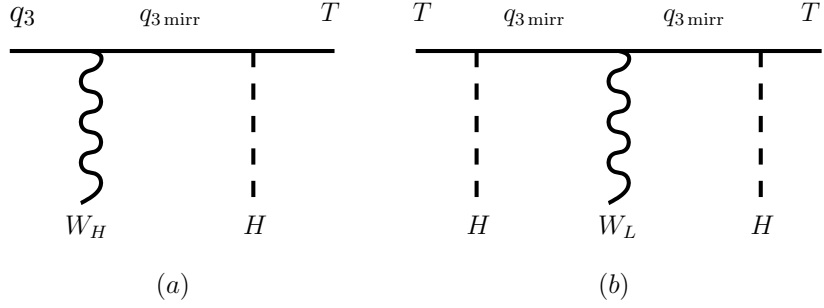}
\caption{\label{IIandIII}{\it Generation of type ${\rm (II)}^\prime$ and
 ${\rm (III)}^\prime$ vertices by integrating out the mirror fermions.}}
\end{figure}
\begin{eqnarray}
{\rm (I)} &\quad& \frac{v}{f}  \{T^c \Phi^0 t, T^c \Phi^+ b\} \\
{\rm (II)} &\quad& g \frac{v}{f}  \{(T \bar{\sigma}_\mu t)
W_H^{0\,\mu}, (T \bar{\sigma}_\mu b) W_H^{+\,\mu}\} \\
{\rm (III)} &\quad& g\frac{v^2}{f^2}
        (T \bar{\sigma}_\mu T) W_L^{0\,\mu}\\
\label{IV}
{\rm (IV)} &\quad& g^\prime  (T \bar{\sigma}_\mu T) B_L^\mu
\end{eqnarray}
Type (I) vertices arise from expanding the top sector lagrangian
(\ref{topnew}) to order $1/f$, after the Higgs gets a VEV from
electroweak symmetry breaking. Type (II) and (III) vertices can
come from integrating out the mirror fermions as shown in
Fig.~\ref{IIandIII} $(a)$ and $(b)$, which give rises to the
interactions
\begin{eqnarray}
\label{IIp}
{\rm (II)}^\prime &\quad& \frac1{m_{\rm mirr}}
  (T H^\dagger \bar{\sigma}_\mu \tau_a q_3) W_{H}^{a\,\mu} \\
\label{IIIp}
 {\rm (III)}^\prime &\quad& \frac1{m_{\rm mirr}^2} (T
\bar{\sigma}_\mu T)
    (H^\dagger \tau_a H) W_L^{a\,\mu}
\end{eqnarray}
where $\tau_a$'s are the $SU(2)$ generators, and
$m_{\rm mirr}$ is the mass of the mirror fermions.
After electroweak symmetry breaking, we obtain type (II) and (III)
interactions. The mass of the top mirror fermion is parametrically
of the same order as $f$ though it can be much heavier in
practice. There are also vertices similar to (II) with $W_H^0$
replaced by $B_H$ and $g$ replaced by $g'$. Type (IV) vertex is
simply the hypercharge interaction of $T$. More detailed
discussions on the interactions involving $T$-odd fermions can be
found in the Appendix.

\begin{figure}[t]
\includegraphics[scale=0.70,angle=0]{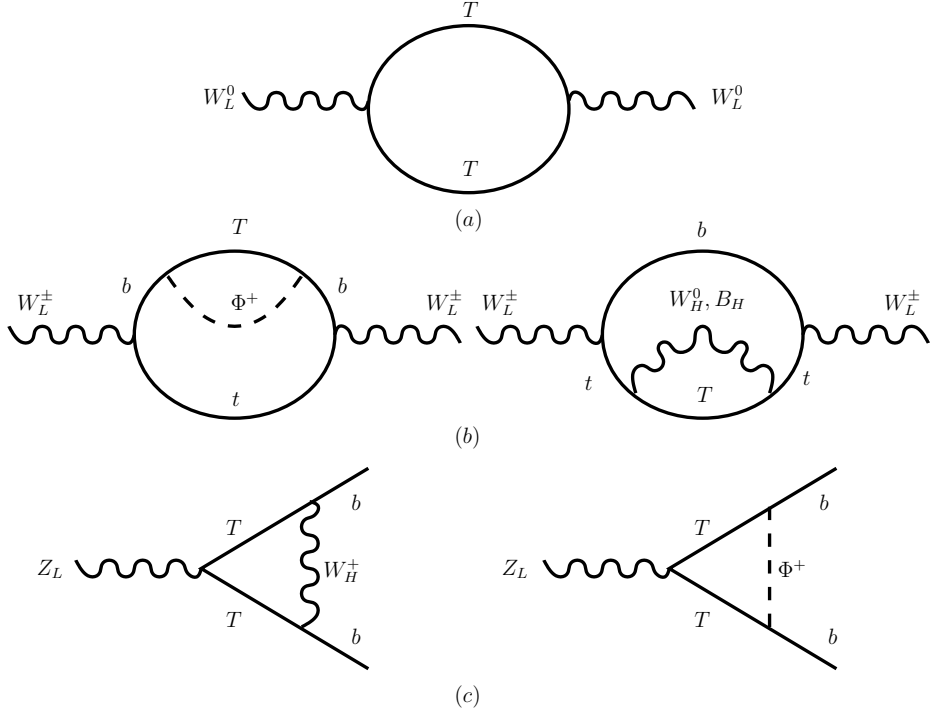}
\caption{\label{figthree}{\it Examples of diagrams contributing to $\rho$ parameters and
$Zb\bar{b}$ vertex.}}
\end{figure}

Let us consider the $\rho$ parameter first. At one-loop order
$(T,T^c)$ contribute to the self-energy of $W_L^0$ through type
(III) vertex, as shown in Fig.~\ref{figthree} $(a)$, and thus
enters into the $\rho$ parameter
\begin{equation}
\delta \rho_T = \frac3{16\pi^2} \frac{e^2}{s_w^2 c_w^2 m_Z^2}
  {\cal O} \left(g^2 \frac{v^4}{f^4}\right) m_T^2 \log
      \frac{m_{\rm mirr}^2}{m_T^2}.
\end{equation}
In the above we have replaced the cutoff dependence in the logarithm with
$m_{\rm mirr}$,  since type (III) vertex arises
from integrating out the mirror fermions. For the littlest Higgs model, this
contribution is parametrically
comparable to the one coming from the heavy gauge bosons~\cite{Cheng:2004yc,Hubisz:2005tx}
\begin{equation}
\label{wcorr}
\delta \rho_{W_H} = -\frac{9}{64\pi^2}\frac{g^2 e^2}{s_w^2 c_w^2 m_Z^2} \Delta m^2
  \log \frac{\Lambda^2}{m_{W_H}^2} ,
\end{equation}
where
\begin{equation}
\Delta m^2 = m_{W_{H}^0}^2 - m_{W_H^{\pm}}^2 =
\frac12 g^2 f^2 \sin^4 \frac{v}{f},
\end{equation}
and $\Lambda=4\pi f \sim 10\ {\rm TeV}$ is the cutoff of the
non-linear sigma model. A partial cancellation is possible between
$\delta \rho_T$ and $\delta \rho_{W_H}$ as they come with opposite
signs. Therefore, the constraint from the calculable contributions
within the low-energy effective theory is expected to be even
weaker than $f\gtrsim 500$ GeV quoted previously in the
$T$-invariant littlest Higgs model with a $T$-even top partner
 \cite{Cheng:2004yc,Hubisz:2005tx}.
Nevertheless, it is still desirable to keep a reasonable hierarchy
between $v$ and $f$ so that the cutoff $\Lambda=4\pi f$ is high
enough to suppress the unknown contributions above the cutoff of
the theory.
There are also two-loop diagrams in $\delta \rho_T$ which involves
type (I) and (II) vertices, two examples of which are shown in
Fig.~\ref{figthree} $(b)$. They are not expected to give
significant contributions.

For the case of $Zb\bar{b}$ vertex, there are now one loop
diagrams using the combinations of the interactions of type
(I)--(IV). Two such diagrams are given in Fig.~\ref{figthree}
$(c)$. These diagrams are suppressed by at least $(v/f)^2$ from
the vertices. Such $(v/f)^2$ suppressions are parametrically of
the same order as the $s_L^2$ factor in the case of a $T$-even top
partner. However, because the exchanged particle is now a heavy
particle such as $\Phi$ or $W_H$, instead of $W_L$ as would be the
case for a $T$-even top partner, there is an additional
suppression factor of $(m_W/m_{\rm heavy})^2$ comparing to the old
expression in Eq.~(\ref{zbb}), which makes the correction to the
$Zb\bar{b}$ vertex even smaller.

\section{LHC Phenomenology}

In this section, we outline the collider phenomenology of this
class of little Higgs models at LHC. We will focus on general
features of the signals and leave more detailed study for the
future.

\noindent{\bf Decay Modes}

Characterization of the decay modes is crucial for identifying
new physics signals. We focus our attention on the following
states \bear T, \ \ A_H,  \ \ W_H. \eear They are the states which
are responsible for canceling the one-loop quadratic divergence
and are generically present in little Higgs models. There should
also be scalars, whose nature is more model-dependent. These
scalars
cancel the quadratic divergence from the Higgs self-coupling and
in general have small production cross-sections at the LHC. Hence
they are more difficult to discover. All other states can
be pushed up to multi-TeV or higher so they may not be relevant at
the LHC. Of course, if some of the extra states happen to be
light, they can give additional signals beyond what is discussed
here.

More specifically, we will narrow our attention further down to
the decay chain of $T$ since it has the largest production
cross-sections. If $W_H$ is involved in the decay chain, its decay
will be important as well.

We summarize various decay channels of $T$ as follows:
\begin{enumerate}
\item $T$ can decay through $T \rightarrow t A_H$, $T\rightarrow t
  W_H^0$, and $T \rightarrow b W_H^+$.
\item Since $T$ is a singlet, it does not have direct renormalizable
  interactions with $W_H$, but
  there can be higher dimensional operators such as $\bar{T} H^\dagger \not{\!\! W_H} q_3$.
  After plugging in the Higgs VEV, the coupling $\bar{T} \not{\!\! W_H} q_3$
  is suppressed at least by $v/f$. This is the type (II) vertex discussed
  in the previous section.
  As a result, the width of $T\rightarrow t
  W_H^0$, and $T \rightarrow b W_H^+$ are suppressed (at least) by
  $v^2/f^2$. Therefore, their branching ratio will be suppressed with
  respect to the dominant mode $T \rightarrow t A_H$ by about
\be
\left( \frac{g}{g'} \right)^2 \frac{v^2}{f^2} \sim 10 \% \left(
  \frac{1 \mbox{TeV}}{f} \right)^2
\ee
\item $T \rightarrow b W_H^+$ with $W_H^+ \rightarrow W^+ A_H$ will
  produce final states identical to $T \rightarrow t A_H$. Therefore,
  they are less distinct.
\item $T\rightarrow t W_H^0$, with $W_H^0 \rightarrow h A_H$ could be
  more interesting. There will be more $b$'s in the final state. In
  addition, since $W_H^0$ decay predominantly to $h A_H$, this decay channel
  could be different from the typical second neutralino decay of
  SUSY; $\tilde{N}_2$ could have sizable branching fractions to both
  $h \tilde{N}_1$ and $ Z \tilde{N}_1$ (in contrast, there is no
  $W_H^0 \to Z A_H$ decay), although the latter is
  somewhat suppressed when $\tilde{N}_1$ is dominantly
  $\tilde{B}$.
\end{enumerate}

\noindent{\bf Production}

We now turn to the discussion of the production of new heavy
states. Analogous to supersymmetry with exact $R$-parity, all new
$T$-odd particles in little Higgs models will have to be pair
produced. This makes their searches at colliders more challenging.
On the other hand, because of the $T$-parity, there is no strong
constraint from the precision electroweak data and the new
particles could be much lighter than those in models without
$T$-parity, which makes them more accessible at the colliders.

Pair productions and decays of the heavy gauge bosons will produce
jets, leptons, and missing energies in the final states, which is
similar to decays of neutralino and chargino productions in
supersymmetry. However, it would be difficult to separate them
from the standard model backgrounds, such as $WZ$ and $WW$.

In addition, the top partner is a Dirac fermion charged under
color. Therefore it should have decent production cross sections.
\begin{figure}[tb]
\includegraphics[scale=0.6,angle=270]{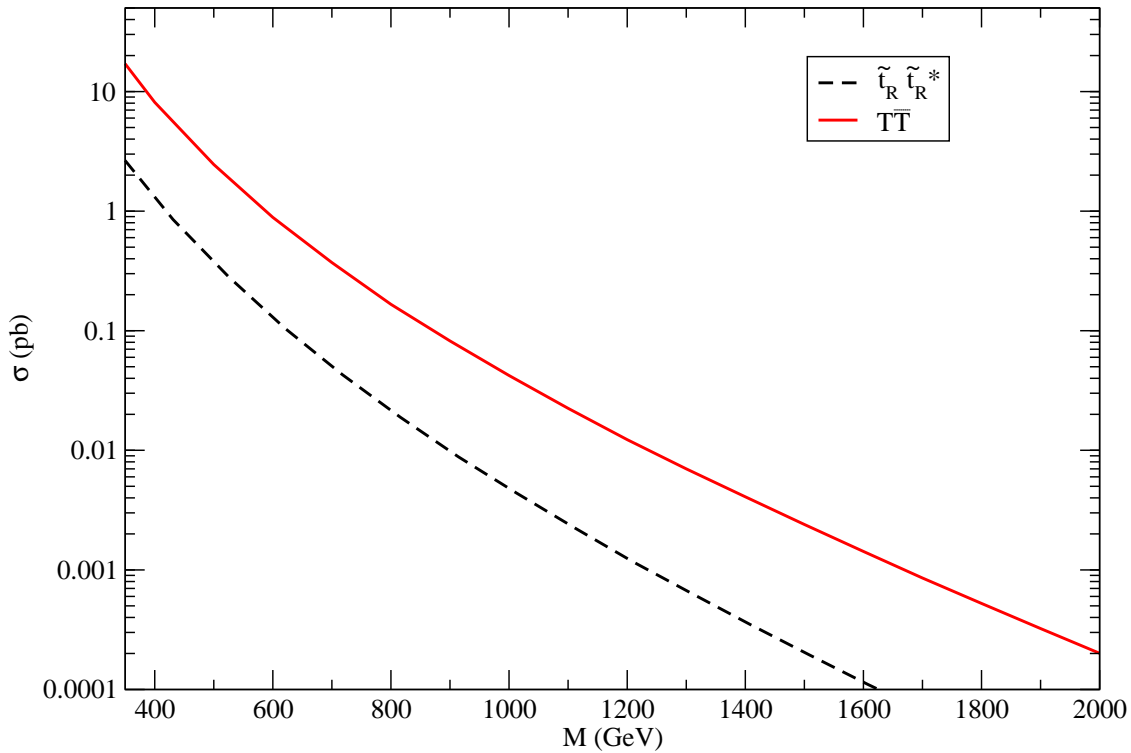}
\caption{\label{tprod}{\it Production cross sections of}\ $T$ {\it
at the LHC as a function of its mass (solid line). For comparison,
we also plot the production cross section of}  \ $\tilde{t}_R$
{\it in SUSY.}}
\end{figure}
\begin{enumerate}
\item As shown in Fig.~\ref{tprod}, the production cross section is
  sizable. Assuming $M_T=1$ TeV, and $L=100$  fb$^{-1}$, thousands of
  $T \bar{T}$ pairs could be produced at the LHC.
\item If the dominant decay channel is $T\rightarrow t A_H$, it will
  be challenging to suppress the top background. A large
  missing energy cut will be necessary and a detailed study is needed to
  tell how well the signals can be separated from the backgrounds.
\item There could be additional decay channels with sizable branching
  fractions. In particular, the decay mode
  $T\rightarrow t W_H^0$ could give us additional signatures.
 \end{enumerate}

\noindent{\bf Comparison with SUSY}

The heavy top partner $T$ cancels the one-loop quadratic
divergence in the Higgs potential coming from the top quark. In
this sense, its role is analogous to the top squark in
supersymmetry. More generally, it is a common feature of models
with a natural mechanism to stabilize the electroweak scale to
have a state with top-like quantum numbers. Such a state could be
copiously produced at the LHC. Its generic signature could be a
set of energetic $b$-rich events above the $t \bar{t}$ background.

The most important task after discovering such a state is
obviously measuring its properties. It is likely to be a
non-trivial task as we will demonstrate with our current example.

We will compare the signatures of $T$ with those of $\tilde{t}_R$
in SUSY. As we commented previously, if the spectrum and the
couplings of the model are such that there are several decay
channels of $T$, there may be some distinctive features by
examining the decay branching ratios of $W_H^0$ as compared with
those of $\tilde{N}_2$ in supersymmetric theories. However, such a
comparison relies on assumptions about the details of the spectrum
and is not generic.
Therefore, we will focus our attention on comparing the case in
which $T$ predominantly decays to $t\, A_H$ with the case in
which $\tilde{t}_R \rightarrow t \tilde{B}$.

\begin{figure}[tb]
\includegraphics[scale=0.6,angle=270]{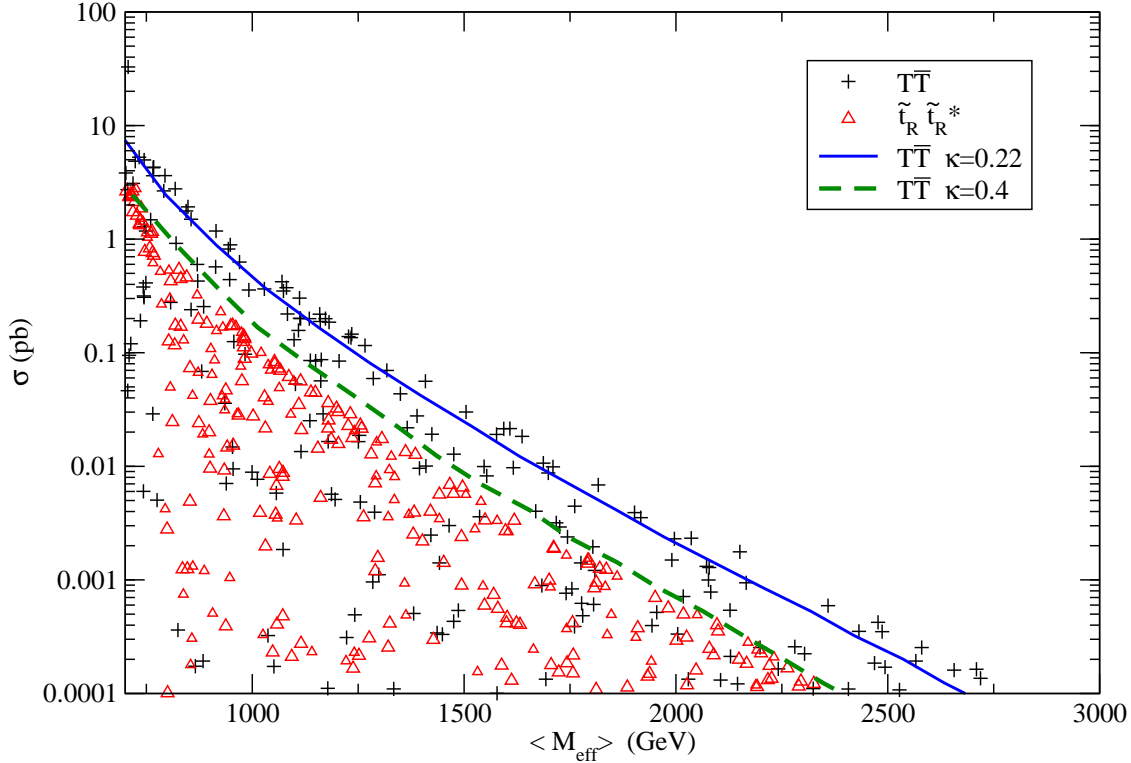}
\caption{\label{sig_meff}{\it Comparison between} $T$ {\it and}
  $\tilde{t}_R$. $\kappa=m_{A_H}/m_T$ or
  $\kappa=m_{\rm LSP}/m_{\tilde{t}_R}$. }
\end{figure}
Since fermions have very different cross sections comparing to
bosons, they might be distinguishable from bosons based on their
production rate\footnote{We assume it is possible to separate a
  significant portion of the signal from the $t \bar{t}$
  background. It is obviously a crucial assumption which deserves
  further study. Although it will not
change the qualitative feature outlined here, inclusion of the effect
of the background will certainly make such a
comparison even more difficult than what is illustrated in this
study.}. However, in order to make a meaningful comparison, one must
get a handle on the mass of new
particles. One possible
observable is the average effective mass. We define the effective mass by
\be
M_{\rm {eff}}=\not{\!\!E}_T+\sum_i |P_T^i|\,(\mbox{Leading jets, } \#
\mbox{ of jets}\ge 2, \ \not{\!\!E}_T \ge 200 \mbox{ GeV}).
\ee
We use the average of the effective mass as an example of a potential
observable related to relevant mass scales in the problem.

We plot the cross sections versus the average effective mass in
Fig.~\ref{sig_meff}. We see that
 for a significant portion of parameter space when there is a
  large separation between $m_T$ and $m_{A_H}$, $\kappa=m_{A_H}/m_T < 0.4$, these two
  cases are distinguishable. The ratio $\kappa$ is model-dependent.
For the littlest Higgs model presented in the Appendix,
$\kappa=\sqrt{2/5} (g'/\lambda_t) \approx 0.22$, so it can be distinguished.

For larger $\kappa$, the decay products are softer. There can be a
SUSY model giving the same values for the two observables, the
total cross sections and the effective mass, so they are not
distinguishable in this strategy. Such an effect persists even as we use
other generic $P_T$-like observables, which is anticipated since most of
those observables only measure mass difference.  This is a
  general lesson for this kind of comparison based on the total
  cross section in combination with this class of observables proportional
to mass  difference between the heavy top-like state and the lightest stable
  neutral state. Therefore, it is difficult to
  use them to completely break the degeneracy.

On the other hand, it is a degeneracy between two
  specific models with very distinct mass spectra: the SUSY model
  has a much lighter spectrum comparing to the little Higgs model.
  Therefore, some kinematic
  distributions, such as forward jets, or the opening angle of the
 $b$-lepton pair
  might help to distinguish these two scenarios. This requires more
detailed studies in the future. Of
  course, spin-correlations will also be very useful in principle to distinguish
  these two scenarios
  \cite{Barr:2004ze,Smillie:2005ar,Datta:2005zs}, which also
  deserves further studies.

\section{conclusion}

In this paper we have shown that there is a class of little Higgs
models in which every new particle responsible for canceling the
one-loop quadratic divergence, including the heavy top partner, is
odd under $T$-parity. Such a construction largely simplifies the
implementation of $T$-parity in the top sector. We discussed the
generic form of the top sector lagrangian along with a comparison
with earlier constructions. A couple of explicit examples of little
Higgs models with $T$-odd top partners are demonstrated in the
Appendix.

The odd $T$-parity  of the top partner completely changes its collider
phenomenology. The main phenomenological features of such a $T$-odd top
partner is outlined in this paper. Due to the conserved $T$-parity, such states have
to be pair produced. On the other hand, for the same reason, the top
partner could be much lighter without violating the constraints from
the electroweak precision tests, and therefore could have a decent
production rate at the LHC.

Generically, the collider signature for such a state is similar to
other TeV scale new physics models with a $Z_2$ parity, such as
low energy supersymmetry. Such a discrete symmetry is also
independently motivated by the dark matter in the universe, which
could be composed of the lightest particle charged under this symmetry.
Therefore, this kind of models provides an interesting playground
for comparison with supersymmetry. We discuss the challenge of
distinguishing these models from supersymmetry at hadron
colliders. The parameter spaces of these two classes of models
which produce similar collider signature generically have quite
different overall mass spectra. This difference points to a
possible direction of distinguishing them by using more detailed
kinematic distributions. Further detailed studies are certainly
required to check the viability of this strategy at the LHC.

\begin{acknowledgments}
This work is supported in part by the National Science Foundation
under grant PHY-0244821 and the Department of Energy under grant
DE-FG02-90ER40542. We acknowledge the hospitality of the Aspen
Center for Physics where part of this work was completed. We would
also like to thank Jay Hubisz and Patrick Meade for pointing out a
mistake in an early version of this paper.
\end{acknowledgments}

\appendix*

\section{Explicit models}
\label{models}

In Section \ref{toppartner} we demonstrated using a simple toy model the general
idea of
constructing a top Yukawa sector in which the top partner is $T$-odd.
In this Appendix we show how to apply the general idea to
the minimal moose and the $SU(5)/SO(5)$
littlest Higgs models with $T$-parity. We will focus on
implementing $T$-parity in the fermionic sector.
For $T$-parity invariant gauge and scalar sectors  we
refer the readers to Refs.~\cite{Cheng:2003ju,Cheng:2004yc,Low:2004xc}.

First we start with the minimal moose model in Ref.~\cite{Arkani-Hamed:2002qx}.
The model consists of two sites
with four copies of $SU(3)$ global symmetry on each site\footnote{This model does not have a custodial
$SU(2)$ symmetry and contains a large contribution to the $\Delta \rho$ parameter
in the nonlinear sigma model lagrangian ~\cite{Kilic:2003mq}, which can easily be
cured by replacing $SU(3)$ with a group which contains a custodial $SU(2)$
such as $SO(5)$ \cite{Chang:2003un}.}, broken down to the diagonal $SU(3)$'s by
four link fields $\Sigma_i \to L_i \Sigma_i R^\dagger_i,\, i=1,2,3,4$. Within each site
an $SU(2)\times U(1)$ subgroup is weakly gauged with equal strength, the diagonal
of which remains unbroken and is identified with the standard model $SU(2)_W\times U(1)_Y$.
In the fermionic sector we introduce two copies of electroweak doublet
fermions, charged under the $L$- and $R$-site respectively,
\begin{equation}
\label{qab}
Q_a = \left( \begin{array}{c}
                  d_a \\
                  u_a \\
                  0
             \end{array} \right)  \to L_i\ Q_a, \quad
Q_b = \left( \begin{array}{c}
                  d_b \\
                  u_b \\
                  0
             \end{array} \right)  \to R_i\ Q_b.
\end{equation}
Under $T$-parity they transform as $Q_a \leftrightarrow Q_b$. Both
$Q_a$ and $Q_b$ carry additional $U(1)$ gauge charges on top of
the ones corresponding to the $T_8$ generators of $SU(3)$'s in
order to have the correct hypercharges. The $T$-even linear
combination $(Q_a + Q_b)/\sqrt{2}$ will be taken to be the
standard model fermions. To make the other combination massive, we
introduce a conjugate doublet transforming
 non-linearly by way of CCWZ \cite{Coleman:sm,Callan:sn}
\begin{equation}
\label{qcc}
Q_c^c = \left( \begin{array}{c}
                  d_c^c \\
                  u_c^c \\
                  0
             \end{array} \right)  \to U_i\ Q_c^c,
\end{equation}
where the matrix $U_i$ non-linearly realizes the $SU(3)$ global
symmetry through $\xi_i \to L_i \xi _i U_i^\dagger = U_i \xi_i
R^\dagger _i,\ \xi_i^2 = \Sigma_i$
\cite{Cheng:2003ju,Cheng:2004yc,Low:2004xc}. Moreover, $\xi_i \to
\Omega \xi_i^\dagger \Omega$ under $T$-parity, whereas the
conjugate doublet transforms as $Q_c^c \to -  \Omega\,Q_c^c$,
where $\Omega={\rm diag}(1,1,-1)$. We can write down the following
$T$-invariant Yukawa-type interactions
\begin{equation}
\label{toddmass} \kappa f \left(\bar{Q}_a\, \xi_2\, Q_c^c -
\bar{Q}_b\,\Omega\, \xi_2^\dagger\,Q_c^c\right),
\end{equation}
where the coefficient $\kappa$ in general is different for
different generations. Eq.~(\ref{toddmass}) gives a mass to the
linear combination $Q_a - \Omega\, Q_b$, that is the
 $T$-odd Dirac-pair $q_{\rm mirr}=
\left(q_c^c, (q_a - q_b)/\sqrt{2}\right)$. This is the mirror
fermion we refer to in Section III. The construction described
here avoids large contributions to the four-fermion
operators~\cite{Low:2004xc}, which would arise from the CCWZ
kinetic terms if the standard model fermions are also realized
non-linearly as in (\ref{qcc}). With the present construction, the
standard model fermions now have standard kinetic terms without
involving any Goldstone bosons, hence giving no large
contributions to the four-fermion operators. Although strong
constraints on four-fermion operators exist only for the light
generations and the lepton sector, we use the same construction
for the top sector for consistency.

Next we consider the top sector, for which
we introduce complete $SU(3)$ triplets of quarks,
\begin{equation}
Q_{3a}= \left( \begin{array}{c}
                    b_{a} \\
                     t_{a} \\
                     U_{a}
               \end{array} \right), \quad
Q_{3b}= \left( \begin{array}{c}
                    b_{b} \\
                     t_{b} \\
                     U_{b}
                \end{array} \right) , \quad
Q_{3c}^c= \left( \begin{array}{c}
                    b_{c}^c \\
                     t_{c}^c \\
                     U_{c}^c
                 \end{array} \right)
\end{equation}
Under $T$-parity, $Q_{3a} \leftrightarrow  Q_{3b}$ and $Q_{3c}^c
\to -\Omega\,Q_{3c}^c$. In particular, $U_a \leftrightarrow U_b$
and $U_c^c \to U_{c}^c$.

The mirror triplet $Q^{c}_{3c}$ marries a linear combination
$Q_{3a}- \Omega\,Q_{3b}$ and become massive through the
interaction in Eq.~(\ref{toddmass}). Note that there is a massive
$T$-even state consisting of $U_c^c$ and $U_a + U_b$. However, it
can be made very heavy without affecting the cancellation of the
quadratic divergence from the top loop. We assume that it is much
heavier than the $T$-odd top partner discussed below. To complete
the spectrum we add two $SU(2)$ singlets $U^{c}_{a}$, $U^{c}_{b}$,
with the $T$-parity transformation rule $U^c_a\leftrightarrow
U_b^c$, as discussed in Sec.~\ref{toppartner}.
There is some freedom in the $U(1)$
charge assignments. One convenient choice was given in Ref.~\cite{Cheng:2003ju}
which we list here in Table~\ref{u1charges}.
\begin{table}[t]
\caption{\label{u1charges} \it The $U(1)$ charge assignments for
$Q_{a}, Q_{b}, U^{c}_{a}, U^{c}_{b}$ fermions. $Q^{c}_{c}$
is not listed because it transforms nonlinearly.
The physical $U(1)_Y$ hypercharge is the sum of the $U(1)$ charges.}
\begin{ruledtabular}
\begin{tabular}[b]{c|cccccc}
  &\, $q_{a}$\, & \, $U_{a}$ \, & \,$U^{c}_{a}$\,
  &\, $q_{b}$\, & \, $U_{b}$ \, & \,$U^{c}_{b}$\,
\\
\hline
  $U(1)_a$ & $\frac1{12}$ & $\frac7{12}$ & $-\frac7{12}$ & $\frac1{12}$ & $\frac1{12}$ &
  $-\frac1{12}$   \\
\hline
  $U(1)_b$ & $\frac1{12}$ & $\frac1{12}$ & $-\frac1{12}$ & $\frac1{12}$ & $\frac7{12}$ &
  $-\frac7{12}$
\end{tabular}
\end{ruledtabular}
\end{table}
The top Yukawa coupling can be written as
\begin{equation}
{\cal L}_t = \lambda f\, Q_{3a}\; \Sigma_1^\dagger\
\left( \begin{array}{c}
                  0  \\
                  0  \\
                  U_b^c
                  \end{array} \right)
+ \lambda f\, Q_{3b} \;\Omega\, \Sigma_1\Omega\,
\left( \begin{array}{c}
                  0  \\
                  0  \\
                  U_a^c
                  \end{array} \right)
 + {\rm h.c.} \quad ,
\label{minitop}
\end{equation}
which in fact introduces mixing between the light $T$-even and the
heavy $T$-even fermions. This mixing can be minimized by taking
the mass of the heavy $T$-even fermions to be around the cutoff
scale $\sim$ 10 TeV. In other words, the heavy $T$-even fermions
can be decoupled from the spectrum below the cutoff without
affecting the cancellation of top quadratic divergences.
Alternatively, if we add additional interactions in the top sector
as follows
\begin{equation}
{\cal L}_t^\prime =  \lambda^\prime f Q_{3b}\Omega \Sigma_2 \Sigma_1^\dagger
   \left( \begin{array}{c}
                  0  \\
                  0  \\
                  U_b^c
                  \end{array} \right)
         +\lambda^\prime f Q_{3a}  \Sigma_2^\dagger\, \Sigma_1 \Omega \,
         \left( \begin{array}{c}
                  0  \\
                  0  \\
                  U_a^c
                  \end{array} \right) + {\rm h.c.} ,
\end{equation}
then the mixing is suppressed when $\lambda^\prime$ is close to
$\lambda$. In fact, the mixing is completely eliminated in the
limit $\lambda^\prime = \lambda$, since when $\Sigma_2=\langle
\Sigma_2 \rangle=\openone$
\begin{equation}
{\cal L}_t+{\cal L}_t^\prime= \lambda f (Q_{3a}+\Omega Q_{3b}) \Sigma_1^\dagger
\left( \begin{array}{c}
                  0  \\
                  0  \\
                  U_b^c
                  \end{array} \right)
      +\lambda f (Q_{3a}+\Omega Q_{3b})  \Sigma_1 \Omega\,
      \left( \begin{array}{c}
                  0  \\
                  0  \\
                  U_a^c
                  \end{array} \right) + {\rm h.c.}
\end{equation}
which is reminiscent of (\ref{topnew}). The $\Sigma_2$ field is inserted to ensure gauge invariance.
When expanding ${\cal L}_t+{\cal L}_t^\prime$ to quadratic order in the Higgs field which resides in
$\Sigma$'s, we
obtain Eq.~(\ref{newexpand}) with the following
$T$-even top quarks  and $T$-odd top partners:
\begin{eqnarray}
\label{tT}
(t,t^c)&=&\left(\ (t_a+t_b)/\sqrt{2},\ (U_a^c+U_b^c)/\sqrt{2}\ \right), \nonumber \\
(T,T^c)&=&\left(\ (U_{a} -U_{b})/\sqrt{2},\ (U^c_{a}-U^c_{b})/\sqrt{2}\
\right).
\end{eqnarray}

The mass of the $T$-odd heavy partner plays an important role in
collider phenomenology. In this model there are two light Higgs
doublets, so the Yukawa coupling $\lambda$ of the top sector
depends on the ratio of the Higgs VEVs. In addition, $m_{W_H}$ and
$m_{A_H}$ receives contributions from VEVs of all link fields
$\Sigma_i$ while $m_T$ does not. Therefore, the ratio between the
masses of the top partner $T$ and the heavy gauge bosons $W_H$,
$A_H$ is not completely fixed, but depends on model parameters.

At this stage it is worth pointing out that to the first order in
the physical Higgs field, Eq.~(\ref{minitop}) contains the vertex
$T^c H q_{\rm 3\, mirr}$, which was used in Fig.~\ref{IIandIII}
$(a)$ and $(b)$ to obtain the interactions (II)$^\prime$ and
(III)$^\prime$ in Eqs.~(\ref{IIp}) and (\ref{IIIp}) after the
mirror fermions are integrated out. On the other hand, the kinetic
terms for the fermions,
\begin{equation}
\bar{Q}_a\, \bar{\sigma}_\mu D^\mu_a\, Q_a + \bar{Q}_b\,
\bar{\sigma}_\mu D^\mu_b\, Q_b ,
\end{equation}
when expressed in terms of the mass eigenstates give rise to the
vertices $g (q_3 \bar{\sigma}_\mu q_{\rm 3\,mirr}) W_H^\mu$ and
$g^\prime (T\bar{\sigma}_\mu T) B_L^\mu$. The former was used
again in Fig.~\ref{IIandIII} $(a)$, whereas the latter is the
type (IV) interaction in Eq.~(\ref{IV}).

Next we turn to the littlest Higgs model which is based on the coset space $SU(5)/SO(5)$.
The non-linear
sigma model field has a vacuum expectation value
\begin{equation}
\langle \Sigma \rangle \equiv \Sigma_0 = \left( \begin{array}{ccc}
                \phantom{aaa}     &      & \openone \\
                     &   1  &          \\
           \openone  &      & \phantom{aaa}
                  \end{array}     \right),
\end{equation}
where $\Sigma$ is a (${\bf 5}\times {\bf 5}$) symmetric tensor
under $SU(5)$: $\Sigma \to V \Sigma V^T, V\in SU(5)$. Again we
introduce two copies of fermionic doublets, embedded in incomplete
$SU(5)$ multiplets, as well as the conjugate fermions which
transform linearly only under the unbroken $SO(5)$:
\begin{equation}
Q_a = \left( \begin{array}{c}
                    q_a \\
                     0  \\
                     0        \end{array} \right) \to V Q_a , \quad
Q_b = \left( \begin{array}{c}
                     0 \\
                     0  \\
                     q_b        \end{array} \right) \to V^* Q_b, \quad
Q_c^c = \left( \begin{array}{c}
                    q_c^c \\
                     \chi^c  \\
                      \tilde{q}_c^c    \end{array} \right) \to U Q_c^c,
\end{equation}
where $U$ again realizes the $SU(5)$ symmetry non-linearly through
$\Sigma=\xi^2 \Sigma_0,\ \xi\to U \xi \Sigma_0 V^T \Sigma_0 = V
\xi U^\dagger $ \cite{Cheng:2003ju,Cheng:2004yc,Low:2004xc}. The
action of $T$-parity on the fermions takes $Q_a \leftrightarrow
\Sigma_0 Q_b$ and $Q_c^c\to -\Omega_5\, Q_c^c$, with $\Omega_5 =
{\rm diag}(1,1,-1,1,1)$. The $T$-odd combination of doublets is
$(q_a-q_b)/\sqrt{2}$ and gets a mass through
\begin{equation}
\label{so5mirr} \kappa f \left( {Q}_a \xi Q_c^c - Q_b \Sigma_0
\Omega_5\, \xi^\dagger Q_c^c \right).
\end{equation}
The above equation is $T$-invariant since $\xi \to \Omega_5
\xi^\dagger \Omega_5$. The $T$-odd massive mirror fermions are
$q_{\rm mirr}=(q_c^c,(q_a-q_b)/\sqrt{2})$, whereas the remaining
fermions $\chi^c$ and $\tilde{q}_c^c$ can be given Dirac masses by
introducing additional particles
\cite{Cheng:2003ju,Cheng:2004yc,Low:2004xc}.

In the top sector we introduce additional singlets as follows
\begin{equation}
Q_{3a} = \left( \begin{array}{c}
                        q_{3a} \\
                           U_a \\
                           0
                       \end{array} \right) , \quad
Q_{3b} = \left( \begin{array}{c}
                           0 \\
                           U_b \\
                           q_{3b}
                       \end{array} \right) , \quad
Q_{3c}^c = \left( \begin{array}{c}
                    q_{3c}^c \\
                     \chi_3^c  \\
                      \tilde{q}_{3c}^c    \end{array} \right)      .
\end{equation}
As in the case with the minimal moose model, Eq.~(\ref{so5mirr})
gives a Dirac mass to the $T$-even singlets $(\chi_3^c, U_a+U_b)$.
The massless combinations are $q_{3a}+q_{3b}$ and $U_a-U_b$, and
 have even and odd $T$-parity, respectively.
  The top Yukawa coupling can be written
as, introducing additional singlets $U_a^c\leftrightarrow U_b^c$
under $T$-parity,
\begin{equation}
{\cal L}_t = \frac12 \lambda f \epsilon_{ijk} \epsilon_{xy}
\left( Q_{a\,i} {\Sigma}_{jx} {\Sigma}_{ky} U_a^c + (\Sigma_0 Q_{b})_i
        \tilde{\Sigma}_{jx}\tilde{\Sigma}_{ky} U_b^c \right) +{\rm h.c.} ,
\end{equation}
where $\tilde{\Sigma}=\Sigma_0 \Omega_5 \Sigma^\dagger \Omega_5
\Sigma_0$ is the image of the $\Sigma$ field under $T$-parity.
Moreover, $i,j,k=1,2,3$ and $x,y=4,5$. Again we can
introduce additional interactions to remove the mixing between the light
and heavy $T$-even fermions
\begin{equation}
{\cal L}_t^\prime = \frac12 \lambda^\prime f \epsilon_{lmn} \epsilon_{rs}
\left[ (\Omega_5 Q_{b})_l \Sigma^\prime_{mr}
         \Sigma^\prime_{ns} U_a^c +
     (\Omega_5 \Sigma_0 Q_{a})_l
       \tilde{\Sigma}^\prime_{mr}
   \tilde{\Sigma}^\prime_{ns} U_b^c \right] + {\rm h.c.} ,
\end{equation}
where $\Sigma^\prime=\Omega_5\Sigma^\dagger\Omega_5$ and
$\tilde{\Sigma}^\prime=
\Sigma_0 \Sigma \Sigma_0$ is the $T$-image of $\Sigma^\prime$.
Moreover, $l,m,n=3,4,5$ and $r,s=1,2$ in the above.
After adding ${\cal L}_t^\prime$ to ${\cal L}_t$ and taking
$\lambda^\prime=\lambda$,
the top quarks and $T$-odd top partners have the same expression as in
Eq.~(\ref{tT}) and the interactions of the $T$-odd fermions follow those
discussed in the minimal
moose model.
Contrary to the previous case, the relation between the masses of
the $T$-odd gauge bosons and top partner in this simplest model is fixed, and
their masses are given by
\begin{equation}
M_{W_H} = g\, f, \quad M_{A_H} = \frac{1}{\sqrt{5}}\, g' \, f, \quad
M_T = \frac{1}{\sqrt{2}}\, \lambda_t \, f .
\end{equation}


\end{document}